# Architecting three-dimensional ZnO nanorods

Argyro N. Giakoumaki,[1,2] George Kenanakis,[1] Argyro Klini,[1] Zacharias Viskadourakis,[3] Maria Farsari,[1] and Alexandros Selimis[1]

[1]*IESL-FORTH, N. Plastira 100, 70013, Heraklion, Crete, Greece.*

[2]*Department of Chemistry, University of Crete, 70013, Heraklion, Crete, Greece.*

[3]*Crete Center for Quantum Complexity and nanotechnology, Physics Department, University of Crete, 70013, Heraklion, Crete, Greece.*

## Abstract

The fabrication of nanostructures with controlled assembly and architecture is significant for the development of novel nanomaterials-based devices. In this work we demonstrate that laser techniques coupled with low-temperature hydrothermal growth enable complex three-dimensional ZnO nanorods patterning on various types of substrates and geometries. The suggested methodology is based on a procedure involving the 3D scaffold fabrication using Multi-Photon Lithography of a photosensitive material, followed by Zn seeded Aqueous Chemical Growth of ZnO nanorods. 3D, uniformly aligned ZnO nanorods are produced, exhibiting electrical conductance and highly efficient photocatalytic performance, providing a path to applications in a diverse field of technologies.

## Letter

Zinc Oxide (ZnO) is a widely studied metal oxide semiconductor, in the center of extensive research efforts, due to enhanced technological interest in the development of novel electronic, optoelectronic and spintronic materials and devices. It is a very promising material with potential use in different applications, such as gas sensors,[1] transparent electrodes in solar cells,[2,3] phototocatalysts,[4] nanolasers,[5] photoelectrochemical cells for hydrogen generation from water splitting,[6,7] photoluminescent devices,[8,9] and organic light emitting diodes.[10,11] Its useful properties but also the plethora of geometries that can be grown (nanorods–NRs, nano-wires, nanobelts, nanosprings, hierarchical nanostructures etc.)[2,12] make it one of the most studied materials in nanoscience. Moreover, ZnO can be doped with metals that can influence or even tailor its optical, electrical or mechanical properties.[6]

For the fabrication of neat and doped ZnO nanostructures a variety of chemical and physical synthesis methods have been adopted, including vapor-liquid-solid (VLS) method,[5,13] chemical vapor deposition (CVD),[14] thermal evaporation,[15] electrochemical deposition in porous membranes,[16] and aqueous chemical growth

(ACG).[17] Most of these growth techniques have been implemented to control the distribution of ZnO nanostructures on various types of substrates such as glass, Si wafers, flexible organic films and wires.[18] The patterning of ZnO nanostructures has been demonstrated mainly on flat surfaces but recently, a new method for the deposition of aligned ZnO nanorods was introduced by some of the authors that is consistent with different types of substrates including those with cylindrical geometry (optical fibre).[1,19]

Here, we demonstrate a innovative method for the fabrication of conducting, fully three-dimensional (3D) ZnO nanorods-coated structures with highly resolved periodicity and their characteristics regarding morphological, electrical and photocalytic properties. The proposed fabrication scheme involves the seeded hydrothermal growth of ZnO NRs on a 3D scaffold of an organic-inorganic hybrid material (SZ2080), fabricated by Multi-Photon Lithography (MPL).[6] The growth of ZnO NRs is based on a two-step procedure that requires the deposition of a metallic zinc (Zn) precursor layer onto the polymeric scaffold, employing pulsed laser deposition (PLD) technique, followed by an aqueous chemical growth of ZnO nanocrystalline rods out of an aqueous solution of zinc nitrate hexahydrate ($Zn(NO_3)_2$) in the presence of ammonia.[20]

The methodology presented is a straightforward and flexible scheme carried out at relatively low temperature (<100º C), and is consistent with different type of substrates. Additionally, the laser-based techniques employed for the fabrication of 3D scaffold (MPL) and the precursor Zn layer (PLD), allow the deposition of ZnO NRs in the form of micro-architected patterns on substrates with flat or complex geometry. Additional variability in the NR structure architecture can be introduced by varying the growth conditions (temperature, growth time etc.), while the incorporation of appropriate chemicals would enable the growth of doped or functionalized ZnO NRs.[21]

In that context, we present ZnO nanostructured 3D architectures prepared with the proposed methodology, and we demonstrate the essential role of the Zn precursor layer for the uniform distribution of ZnO NRs over the surfaces of the assembly. The properties of these structures, regarding morphology, structure and conductivity have been examined. Additionally, their photocatalytic efficiency has been investigated, and is found to increase dramatically compared to the efficiency of ZnO NRs of similar characteristics that has been deposited on flat substrates.

A detailed description of the experimental techniques, set-ups, conditions and materials employed is provided in the supplement.

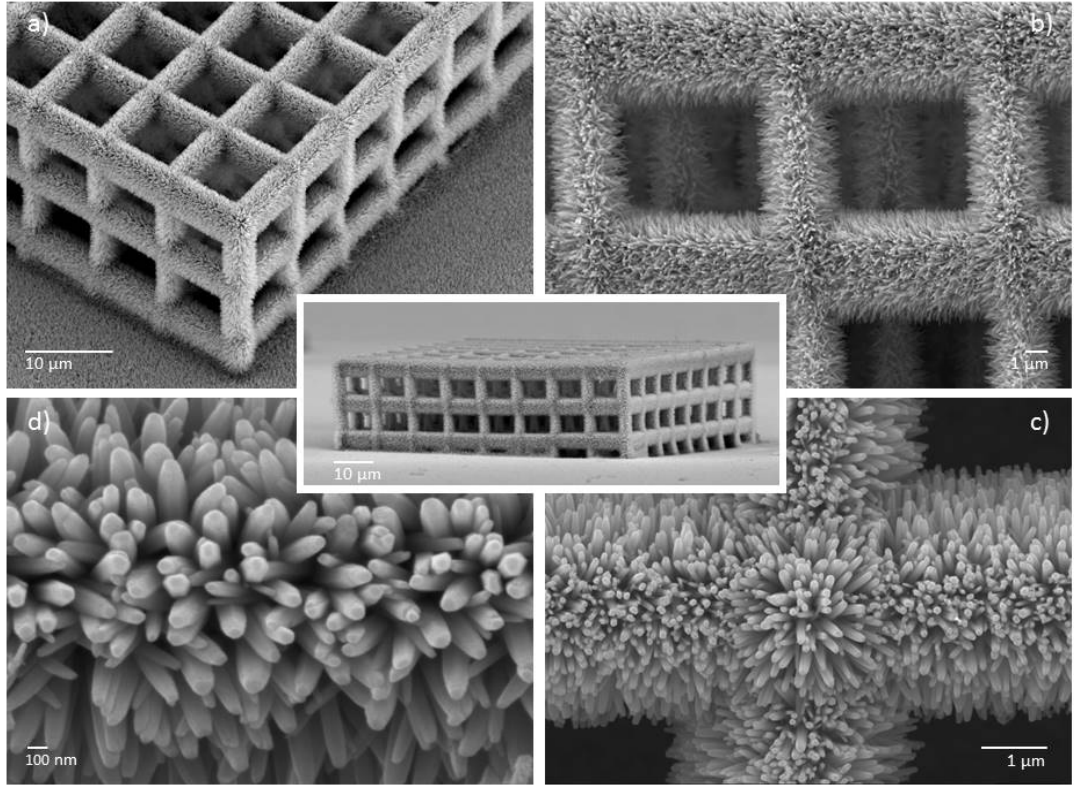

**Figure 1: SEM images of a ZnO NRs coated 3D structure of blocks (center), fabricated with MPL and Zn seeded ACG. The scaffold has been fabricated with MPL galvanometric scanner system that is described in the supplement (laser power=10 mW, objective: 100x, scanning speed=20 μm/s). ACG growth time: 2 h**

Figure 1 shows a Scanning Electron Microscope (SEM) picture of a ZnO NRs coated 3D periodic structure, grown via the proposed fabrication scheme. It consists of parallel blocks of the same dimensions (10x10x10 μm$^3$). The SEM images confirm the top-to-bottom uniformity and long-range continuity of the ZnO nanorod layer over the surfaces (Figure1a and c), even at the interior parts of the 3D assembly (Figure 1b). The structure is covered by dense well-aligned ZnO nanorods, having diameter and length of 50 nm and 1 μm, respectively (Figure 1d).

In order to investigate the role of the Zn seed layer in the quality of ZnO NRs patterning, a number of pristine, without the Zn precursor layer, 3D scaffolds were chemically treated in the growth solution to synthesize ZnO nanorods. Figure 2 shows the SEM images of a ZnO NRs covered 3D structure consisting of a number of columns heads (Figure 2a) fabricated with MPL. As evidenced, the grown ZnO features (Figure2b) exhibit different characteristics compared to those grown on Zn coated 3D structures (Figure1) resembling a non-uniform, random growth throughout the deposited area, especially in the vertical side of the columns. Additionally, the formed ZnO NRs are of different shape and dimensions (Figure2c). This is in agreement with a previous study underlying the significant role of the morphology of the Zn seed layer on the characteristics of the ZnO nanostructures grown over it.[22]

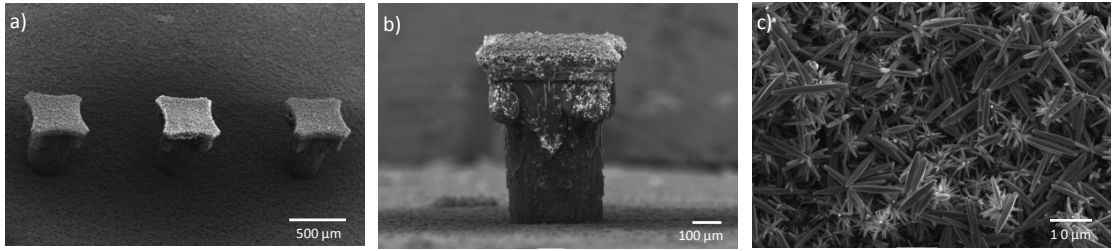

**Figure 2:** SEM images of ZnO NRs chemically grown on a periodic structure of columns, without the deposition of Zn seed layer (a), (b) Top and side view of the columns respectively, (c) ZnO NRs on the top of the column. The scaffold has been fabricated by MPL technique using the galvanometric scanner system described in the supplement (laser power=35 mW, objective: 10x, scanning speed=0.1 mm/s). ACG growth time: 3 h

The photocatalytic activity of the ZnO NRs-coated 3D structures was quantified by measuring the decolourization of methylene blue (MB) in aqueous solution (initial concentration: $5.4 \times 10^{-7}$ mol/L (20 ppm), pH: 5.5). This is a typical potent cationic dye widely used as a model organic probe to test the photocatalytic performance of photocatalysts.[23-25] The samples tested possess the complex geometry presented in Figure 3. The whole sample (0.5 mm x 0.5 mm) consists of an array with 7x7 (49) stacks of circles. Each stack comprises 6 layers of 85 circles, with 50 μm diameter. Figure 3a shows a section of the structure prior ZnO nanorods coating, while in Figure 3b a stack of the structure with a ZnO NRs overlay (Figure 33c) is presented. For comparison, the photocatalytic performance of ZnO NRs, fabricated by the same growth procedure, on SZ2080-coated flat (glass) substrate of the same dimensions (0.5 mm x 0.5 mm) was investigated.

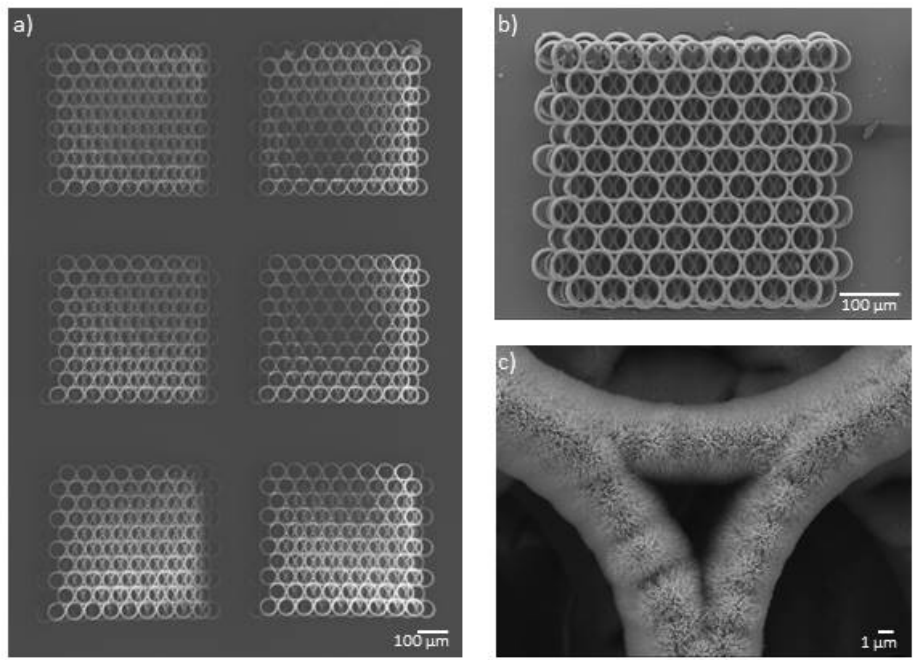

**Figure 3:** SEM images of a ZnO NRs decorated 3D array of circle stacks. (a) section of the 3D array pristine scaffold, (b) a stack of the array, deposited with ZnO NRs, (c) detail of the ZnO NRs deposited on the circles of the stacks. The scaffold has been fabricated with MPL technique using the galvanometric scanner system described in the supplement (laser power=35 mW, objective: 10x, scanning speed=0.1 mm/s). ACG growth time: 2 h

The samples were placed in a custom made quartz cell, and the whole setup was illuminated for up to 60 min using a UV lamp centred at 365 nm (Philips HPK 125 W) with a light intensity of ~10 mW/cm$^2$. The MB concentration (decolourization) was monitored by UV-Vis spectroscopy in absorption mode, at the peak wavelength ($\lambda_{max}$ = 665nm), using a K-MAC spectrophotometer over the wavelength range of 220-800 nm. In addition, the apparent rate constant (k) was calculated as the basic kinetic parameter for the comparison of the photocatalytic activities. The rate constant k was obtained by fitting the concentration data to the equation $\ln(C_t / C_0) = -k\,t$, where $k$ is the apparent rate constant, $C_t$ is the actual concentration at time $t$ and $C_0$ the initial concentration of MB. Since methylene blue has a strong absorption peak at 665 nm, a series of MB aqueous solutions of known concentration were prepared first in order to build a calibration curve.

The normalized integrated area of the absorption spectra, at the peak wavelength, of the ZnO nanorods samples, as a function of the UV illumination time is depicted in Figure 4.

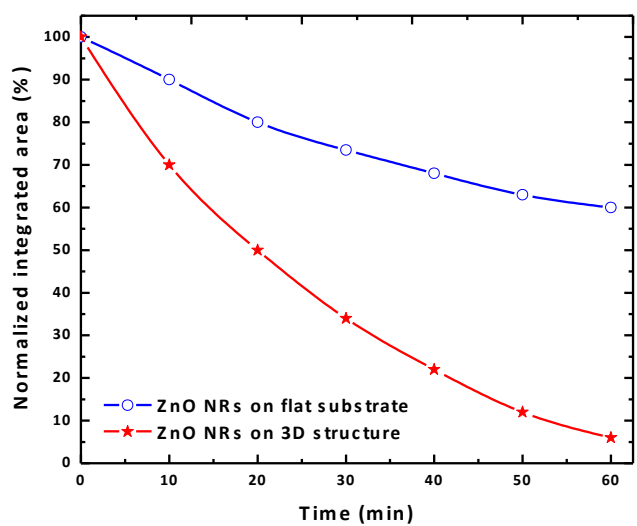

Figure 4: Normalized absorption of aqueous solution of methylene blue at $\lambda_{max}$=665nm, under UV light irradiation for ZnO nanorods grown on the structure of Figure 3 (red stars) and on flat substrates (blue circles).

As it can be seen the ZnO NRs coated 3D structure (stars) exhibit high photocatalytic activity, when compared to the activity of the ZnO NRs grown on flat substrate (circles). Particularly, the degradation of MB for the ZnO NRs on the 3D MPL written structure is about 95% in only 60 min (even ~75% in only 30 min), while for the ZnO nanarods on flat substrate the degradation is only ~35% after a period of 60 min. This is in agreement with previous results reported in the literature[26,27] and can be attributed to the high surface-to-volume-ratio of the 3D structures, enhanced further with ZnO NRs deposition. The good linear fit of equation $\ln(C_t / C_0) = -k\,t$ confirms that the photodegradation of MB on ZnO follows first-order kinetics. The apparent rate constant for the ZnO NRs coated 3D laser-written samples was found to be 0.046 min$^{-1}$, which is higher than the reported values for ZnO coated hierarchically roughened silicon surfaces.[24] The calculated rate constant approaches the values reported for $TiO_2$ thin films and nanostructures, which are considered as the most efficient photocatalysts.[25] It worth to note that our reference samples (pristine glass substrate and 3D structures) were also characterized, in terms of MB decolorization, by means of UV-Vis absorption and their photocatalytic activity proved to be negligible.

To summarize, in this letter we have introduced an innovative approach for the reproducible fabrication of complex 3D ZnO NR structures, via combining MPL, a laser based direct writing technique, and a low temperature chemical growth (ACG) which is seeded by a Zn layer deposited by PLD technique. The grown structures exhibit excellent ZnO NRs covering and vertical alignment over the surfaces of the assemblies, as well as crystalline structure and electrical conductivity. We have also shown that the transition from 2D to 3D architectures results to a significant increase of the photocatalytic ability of the material.

The key motivation of our work presented herein was to illustrate the principles of the fabrication scheme for the development of ZnO based 3D structures, with excellent morphological characteristics and physical properties. Secondly, we believe that the many-fold enhanced active area of these new nanostructures allows for their integration in different technological fields and devices[6,28-32], such as gas sensing, photo-detecting, phocatalysis or water splitting and we are continuing our investigation with emphasis in that direction.

Moreover, as it was thoroughly discussed, the introduced methodology for the development of 3D nanostructured geometries requires the presence of a uniform thin film over the MPL written scaffold, which is deposited via a well-established, versatile deposition technique, PLD. This provides us with the confidence that the proposed fabrication scheme can be easily extended to a variety of materials and coatings such as metals, oxides, even polymers, as long as the hydrothermal growth temperature doesn't exceed the temperature that 3D hybrid scaffold degrades (180°C), opening the path for new applications in plasmonics and metamaterials.[33]

Finally, the disadvantages of this method are related to the slow speed and high cost of the MPL technique. However, MPL is the only available technique allowing the direct printing of free-form, complex 3D structures with the required resolution. Given the unique technology capabilities, there is a lot of concentrated research trying to improve its productivity by either developing faster and biocompatible photoinitiators, or by developing high-aspect ratio[34] or holographic MPL,[35,36] and this technology is progressing fast from the laboratory to the factory floor. If freeform structures are not needed, MPL is not necessary for the fabrication of 3D nanostructures. Another technique which could provide 3D periodic high-resolution polymeric structures is multi-photon interference lithography.[37] Using this technique, 3D periodic high-resolution structures could be made using only one (or a few) pulses of light; then the proposed methodology could be used to further functionalize the structures with zinc and ZnO.

## Acknowledgements

This research has been co-financed by the European Union (European Social Fund - ESF) and Greek national funds through the Operational Program "Education and Lifelong Learning" of the National Strategic Reference Framework (NSRF) - Research Funding Program: THALES, Projects: na(Z)nowire (MIS 380252) and 3DSET (MIS 380278) as well as by ARISTEIA II Excellence "PHOTOPEPTMAT" program (3941) of the Greek Secretariat for Research and Technology. We would like to thank Profs. A. Andriotis and D. Anglos for may fruitful discussions.

## Author Contributions

AK and MF developed the original concept. AG performed the material synthesis and fabrication experiments. GK carried out the structure characterization, and the photocatalysis experiments and analysis. ZV did the electrical characterization. AS supervised and coordinated the experimental work. All authors contributed to the writing of the manuscript.

Correspondence: Argyro Klini (klini@iesl.forth.gr) and Maria Farsari (mfarsari@iesl.forth.gr)

## Supplementary Section

**Scaffold material**

The photopolymer used for the fabrication of the 3D scaffolds is the organic-inorganic hybrid coded as SZ2080 and is described in reference 38. Its main components are [3-(Methacryloyloxy)propyl] trimethoxysilane (MAPTMS), methacrylic acid (MAA) and Zirconium n-propoxide (ZPO). As photo-initiator (PI) 4,4'-bis (diethylamino) benzophenone was used.

MAPTMS was hydrolyzed using HCl solution (0.1 M) at a ratio of 1:0.1 and ZPO was stabilized by MAA (molar ratio 1:1). After 5 minutes, the Zirconium complex was slowly added to the hydrolyzed MAPTMS at a 2:8 molar ratio. Finally, the PI, at a 1% mass ratio to the monomers was added to the mixture. After stirring for 15 minutes further, the composite was filtered using a 0.22 µm syringe filter.

The samples were prepared by drop-casting onto 100 micron-thick silanized glass substrates, and the resultant films were dried at 100°C for 10 minutes before the photopolymerization. After the completion of the component laser writing process, the samples were developed for 20 minutes in a 30% v/v solution of 4-methyl-2-pentanone in isopropanol and were further rinsed with isopropanol.

**Multiphoton Lithography (MPL)**

The method that was applied for the fabrication of the 3D scaffolds, which serves as substrate for the ZnO NRs growth, is Multiphoton Lithography (MPL) technique.[39,40]

MPL is a laser-based additive manufacturing technique, based on the phenomenon of multiphoton absorption, and allows the direct fabrication of fully 3D microstructures with high resolution. When the beam of ultrafast laser pulses is tightly focused inside the volume of a transparent, photopolymerisable resin, the high intensities within the focused beam voxel can cause the material to absorb more than one photons, resulting in the local photopolymerision. Moving the laser beam inside the material, 3D structures can be directly "written"; all that is needed afterwards is to remove the unexposed, unpolymerized resin, by immersing the sample into an appropriate solvent.

Two different experimental set-ups were employed in this work. The first system, thoroughly described in Ref. [41], offers the fabrication of complex 3D structures at high speed (Figure2 and 3). It comprises of a galvanometric scanner-based system (Scanlabs Hurryscan II, computer-controlled by SCAPS SAMLight software), where the focused laser beam is scanned through the photopolymeric sample to "write" a predetermined design, while the sample motion is restricted to the z-direction. The light source is a Ti:Sapphire femtosecond laser (Femtolasers Fusion, $\lambda$ = 800 nm, 75 MHz, $\tau$ < 20 fs) and is focused into the photopolymerisable composite using focusing microscope objective lenses with high numerical aperture (100x, N.A. = 1.4, Zeiss, Plan Apochromat and 10x, N.A. =0.45, Zeiss, Plan Apochromat). Before entering the scanner, the laser beam was expanded five times (5x) using a telescope lens to illuminate the full back aperture of the microscope objective and to achieve optimal focussing. Z-axis scanning and larger-scale x-y movements were possible by using a high-precision three-axis linear translation stage (Physik Instrumente). Beam on/off and power were further controlled by a mechanical shutter (Uniblitz) and a motorized attenuator (Altechna), respectively. The power used for the fabrication of the structures was 30-100 mW, measured before the objective, while the average transmission was 20%. The scanning speed was 100-20000 μm/s. The direct write process was monitored by a CCD camera mounted behind a dichroic mirror.

The second MPL system[42] employed (Figure 1), provides superior accuracy compared to the first one, but lower speed. It uses the same ultra-fast laser source and microscope objectives. For the fabrication of a polymeric scaffold, the laser beam remains immobile, while the sample is moved using a linear xyz piezoelectric stages system, which allow fine and step movement (Physik Instrumente). The average power applied for the structure fabrication was 5-15 mW, measured before the objective, while the average transmission was 20%. The scanning speed was set to 10-20 $\mu$m/s.

**Pulsed Laser Deposition (PLD)**

The 3D scaffolds fabricated by MPL, were subsequently covered with a Zn thin film, acting as a precursor for the chemical growth of ZnO nanostructures. The method applied for the deposition of Zn layer was PLD technique.[19] Zn films were grown on 3D polymeric scaffolds in high vacuum conditions ($10^{-7}$ mbar) using a standard PLD apparatus described elsewhere.[43] A UV laser beam (KrF excimer laser, $\lambda$ = 248 nm, $\tau$ = 30 ns) was delivering a series of 2000 pulses at a repetion rate of 10 Hz. The beam was incident onto a bulk Zinc rotating target (Zinc foil 99.95%, GoodFellow) at an angle of 45° with respect to the target normal and was focused by a spherical lens (f = + 30 cm) to yield a laser fluence of 3 J/cm$^2$. The ablated material was collected on the 3D scaffolds, placed parallel to the target at a distance of 5cm from the target surface. Depositions on flat substrates under identical conditions have resulted in Zn layers approximately 40 nm thick.

**Aqueous Chemical Growth of ZnO NRs**

Following Zn deposition, the Zn seeded 3D scaffolds were chemically treated in a solution of 100 mL of 0.02 M aqueous solution of zinc nitrate hexahydrate $Zn(NO_3)_2 \cdot 6H_2O$ (Sigma-Aldrich, 99.0%) and 3.1 mL ammonium hydroxide (28% wt $NH_3$ in $H_2O$, Fluka). More specifically, the Zinc-coated 3D structures were immersed vertically in the ammoniac zinc hydrate solution, to avoid deposition of ZnO nanorods due to gravity, under constant stirring at room temperature. The solution was gradually heated up to the temperature of 95°C which kept stable for 2-5 hours. The as grown samples were rinsed with deionized water.

## Characterization

The ZnO NRs structural properties was deduced with X-ray diffraction (XRD), Raman and Fourier-Transform Infra-red (FTIR) spectroscopy for nanostructures deposited on flat glass substrates coated with SZ2080 hybrid. The procedure followed for the ZnO NR growth was identical (Zn seeded ACG growth) with the one applied for growth on 3D scaffolds. The electrical properties of the ZnO NRs were also measured on the same flat specimen.

**Scanning Electron Microscopy**

The Scanning Electron Microscope (SEM) images were recorded using the JEOL JSM-6390 LV model, at an accelerating voltage of 15 kV.

**X-Ray diffraction**

The crystalline structure of the ZnO nanostructures fabricated on SZ2080 coated glass substrate was investigated through X-ray diffraction (XRD) using a Rigaku (RINT 2000) diffractometer with Cu Ka X-rays, at $\theta/2\theta$ configuration, in the $2\theta$ range of 30.00°-70.00°. Figure 5 presents a typical XRD pattern of ZnO nanorods. The pattern indicates a clear preferential growth orientation along the [002] crystallographic direction, i.e. perpendicular to the glass substrates, in agreement with the JCPDS card file No. 36-1451; no characteristic peaks of Zn or other impurities were detected. The diffraction peaks can be indexed to hexagonal wurtzite structure of ZnO crystal with lattice parameters values of α=3.249 Å and c=5.206 Å. As a result, a ratio of c/a=1.602 is deduced, value very close to that for the ideal hexagonal structure (c/a=1.633)[44], indicating a very high crystalline quality of the ZnO structures grown chemically by ACG.

The superior alignment of the ZnO nanorods on Zn seeded substrates is attributed to the spontaneous oxidation of Zinc metal due to the present of oxygen, naturally dissolved in water and was accelerated by the ammonium oxide solution mentioned in the supplement, at 95˚C.

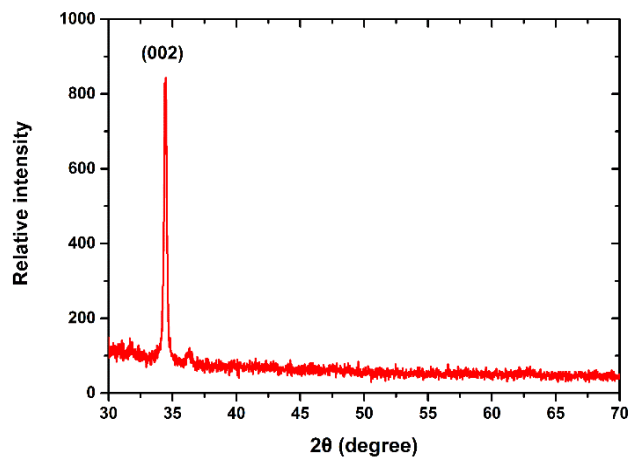

**Figure 5: Typical XRD pattern of ZnO nanostructures synthesized by Zn seeded ACG on SZ2080 coated glass substrates at 95°C.**

### Conductivity Measurements

Electrical conductivity (in S/m) was extracted from impedance measurements. Again, as uniformity was required, samples deposited on flat glass substrates coated with SZ2080, were measured. The procedure followed for the ZnO NR growth was identical (Zn seeded ACG growth) with the one applied for growth on 3D structures.

Figure 6 shows the electrical circuit used to measure the impedance of the ZnO NRs samples.

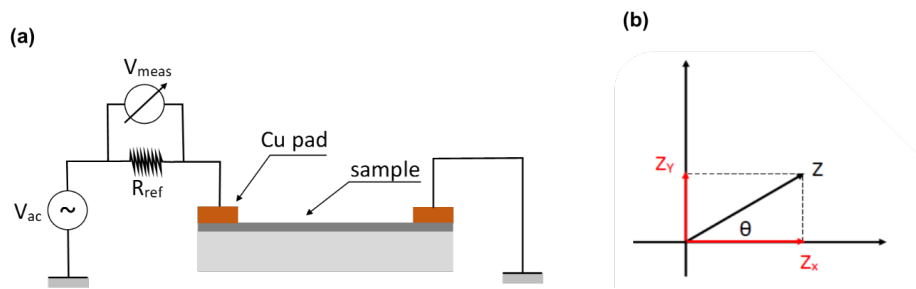

**Figure 6: Electrical circuit used for the impedance measurements (a), and Impedance vector Z and its projections to the real ($Z_x$) and the imaginary axis ($Z_Y$), respectively (b).**

Two Cu pads are mechanically attached to the samples surface. An AC voltage source is connected to one pad through a reference resistance $R_{ref.}$, while the second pad is grounded. A lock-in amplifier is used to measure the voltage drop $V_{meas}$ across the reference resistance and an angle $\theta$ which corresponds to the angle between the impedance vector and the horizontal axis (see Figure 6b). Then the real ($Z_x$) and the imaginary ($Z_Y$) part of the impedance are calculated. $Z_Y$ is correlated to the dielectric permittivity $\varepsilon'$ of the sample, while $Z_x$ corresponds to the dielectric losses and it is related to the conductivity $\sigma$ of the sample through the following relations:

$$\varepsilon' = \frac{V_{meas} \cdot \sin\theta}{2\pi \cdot f \cdot V_{ac} \cdot R_{ref} \cdot \varepsilon_0} \frac{l}{w \cdot t} \qquad (1)$$

$$\sigma = \frac{1}{\rho} = \frac{V_{meas} \cdot \cos\theta}{V_{ac} \cdot R_{ref}} \frac{w \cdot t}{l} \qquad (2)$$

where:

$\rho$: resistivity

$f$: frequency of the ac signal

$\varepsilon_o$: dielectric constant of vacuum $8.85 \times 10^{-12}$ Fm$^{-1}$

$l$: distance between Cu pads

$w$: sample width

$t$: sample thickness

Taking into account the thickness of our samples (~1 μm), we have calculated that the conductivity of the ZnO nanostructured samples is ~0.217 S/m. This metallic behaviour is most likely due to the presence of defects, oxygen vacancies as well as structural defects due to size, shape, and surface effects (polar surfaces causing charge transfer effects).[45] It should be made clear that the conductivity will change as the sample thickness change, and this depends on the NR growth conditions.